\documentclass[rsi,twocolumn,graphicx]{revtex4-1}

\usepackage{graphicx}
\usepackage{dcolumn}
\usepackage{bm}
\usepackage{url}
\usepackage{color}
\usepackage{placeins}
\usepackage{hyphenat}
\usepackage{float}
\usepackage{enumitem}

\usepackage[normalem]{ulem}
\usepackage[colorlinks]{hyperref}

\begin{document}

\title{
Polarization- and time-resolved nonlinear multi-photon spectroscopy for confocal microscopy of semiconductor nanostructures
}

\author{Nikita V. Siverin$^1$, Andreas Farenbruch$^1$, Dmitri R. Yakovlev$^{1}$, Daniel~J.~Gillard$^2$,  Xuerong Hu$^2$, Alexander I. Tartakovskii$^2$, and Manfred Bayer$^{1,3}$}
\affiliation{$^{1}$ Experimentelle Physik 2, Technische Universit\"at Dortmund,                44227 Dortmund, Germany}
\affiliation{$^2$ Department of Physics and Astronomy, University of Sheffield, Sheffield S3 7RH, UK}
\affiliation{$^3$ Research Center FEMS, Technische Universit\"at Dortmund, 44227 Dortmund, Germany}

\date{\today}

\begin{abstract}

We present a versatile confocal microscopy setup for optical second harmonic generation (SHG) and multi-photon spectroscopy that enables polarization-resolved studies of semiconductor bulk crystals and low-dimensional structures. The system offers full polarization control in both excitation and detection, spatial scanning with micrometer resolution, and spectrally tunable excitation over a broad energy range from 0.5 to 4.0\,eV, using femtosecond and picosecond laser pulses. Samples are mounted in a helium-flow cryostat, allowing temperature control from 4 to 300\,K. Magnetic fields up to 0.625\,T can be applied in the Voigt geometry via an electromagnet. The nonlinear optical signals are analyzed using a high-resolution spectrometer with a spectral resolution of 60\,$\mu$eV. We demonstrate the potential of the setup by means of SHG polarization tomography measurements on a Cu$_2$O crystal as well as through a SHG spectral scan of a ZnSe crystal over a wide energy range from 1.4 to 3.1\,eV. Polarization-resolved confocal SHG mapping of various twisted mono- and bilayer MoS$_2$ structures is also presented. In addition, time-resolved two-color pump-probe experiments are shown for a Cs$_2$AgBiBr$_6$ crystal, illustrating the potential of the system for investigating coherent exciton and phonon dynamics.

\end{abstract}


\maketitle 

\section{Introduction}

Nonlinear optical techniques such as second-harmonic generation (SHG) and multi-photon spectroscopy have become powerful tools for investigating material properties that are inaccessible to linear optical methods. In particular, SHG is sensitive to inversion symmetry breaking and can reveal information about the crystal structure, electronic states, and symmetry-related optical selection rules of solids~\cite{Shen,Boyd,Yakovlev18,Fiebig2005,Kirilyuk:05}. While SHG is frequently used non-resonantly, where both the fundamental and second harmonic photon energies are below the band gap~\cite{Spychala2020,Nordlander2018,Aghigh2023-BiophysRev,JamesCampagnola2021}, it becomes especially insightful when performed under resonant conditions, for instance at exciton the resonances in semiconductors. In such cases, the SHG response can be significantly enhanced and reveal rich information on the fundamental properties of solids~\cite{Mund2018-PRB,Mund2019-PRB,Mund20-ZnSeMQW,Farenbruch20,Brunne15,Lafrentz13,Warkentin2018-PRB-THG}.

SHG microscopy (MS-SHG) has become an established tool for probing two-dimensional materials, allowing spatially-resolved mapping of structural properties such as lattice orientation and interlayer coupling ~\cite{Wang2019-OME,Maragkakis2019-OEA,Psilo24review,Zhou2015-JACS,Zeng2013-SciRep}. Most SHG-based microscopy approaches, however, rely on fixed laser wavelengths and analyze only the SHG signal intensity, often without full control or analysis of the light polarization. This limits the ability to extract detailed information about the underlying symmetry and optical selection rules of the investigated materials. One major challenge lies in possible distortion of polarization diagrams caused by the many optical elements present in the setup, which complicates the interpretation of polarization-resolved measurements~\cite{Gusachenko2010-OE,Suceava2025Optica}.
Several approaches have been proposed to compensate for s and p polarization distortions and enable polarization-resolved SHG microscopy ~\cite{Romijn2018-PLOSONE,Chou2008JBO,Kumar2015JBiophotonics}. However, these methods are typically limited to fixed excitation photon energies and lack the flexibility required for broadband or resonant experiments. 

\begin{figure}[b]
	\begin{center}
            \includegraphics[width=\columnwidth]{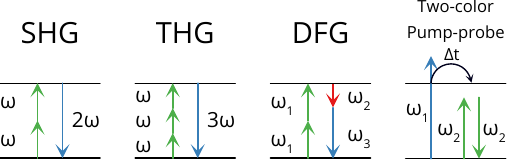}
\caption{Energy level scheme of nonlinear optical processes accessible with the implemented setup: second harmonic generation (SHG), third harmonic generation (THG), difference frequency generation (DFG), and a two-color pump-probe scheme for time-resolved spectroscopy. Arrows indicate the optical transitions in the excitation (upward arrows) and emission (downward arrows) paths.
        }
		\label{fig:setup_schemes}
	\end{center}
\end{figure}

To overcome these limitations, we develop a versatile confocal microscopy setup for optical harmonic generation, including second-, third-, and fourth-harmonic generation (SHG, THG, FHG), as well as for multi-photon spectroscopy techniques such as sum-frequency generation (SFG), difference-frequency generation (DFG), and four-wave mixing (FWM), see Fig.~\ref{fig:setup_schemes}. 
The setup provides full control over the linear polarization direction in both excitation and detection paths, supports a broad tuning range of laser excitation ($0.5-4.0$\,eV), and enables measurements across a wide temperature range ($4-300$\,K), as well as application of an external magnetic field up to 0.625\,T.

The setup is designed to study both bulk crystals and nanostructures, including 2D materials with lateral dimensions of a few micrometers. It supports spatial scanning for SHG mapping, time-resolved pump-probe experiments using synchronized femtosecond and picosecond pulses, and high-resolution spectral detection limited by the spectrometer. In this paper, we describe the technical implementation of the setup and demonstrate its potential on representative test structures, including bulk semiconductors and atomically thin materials.

\section{Technical implementation }
\begin{figure*}[!hbpt] 
	\centering
        \includegraphics[width=\textwidth]{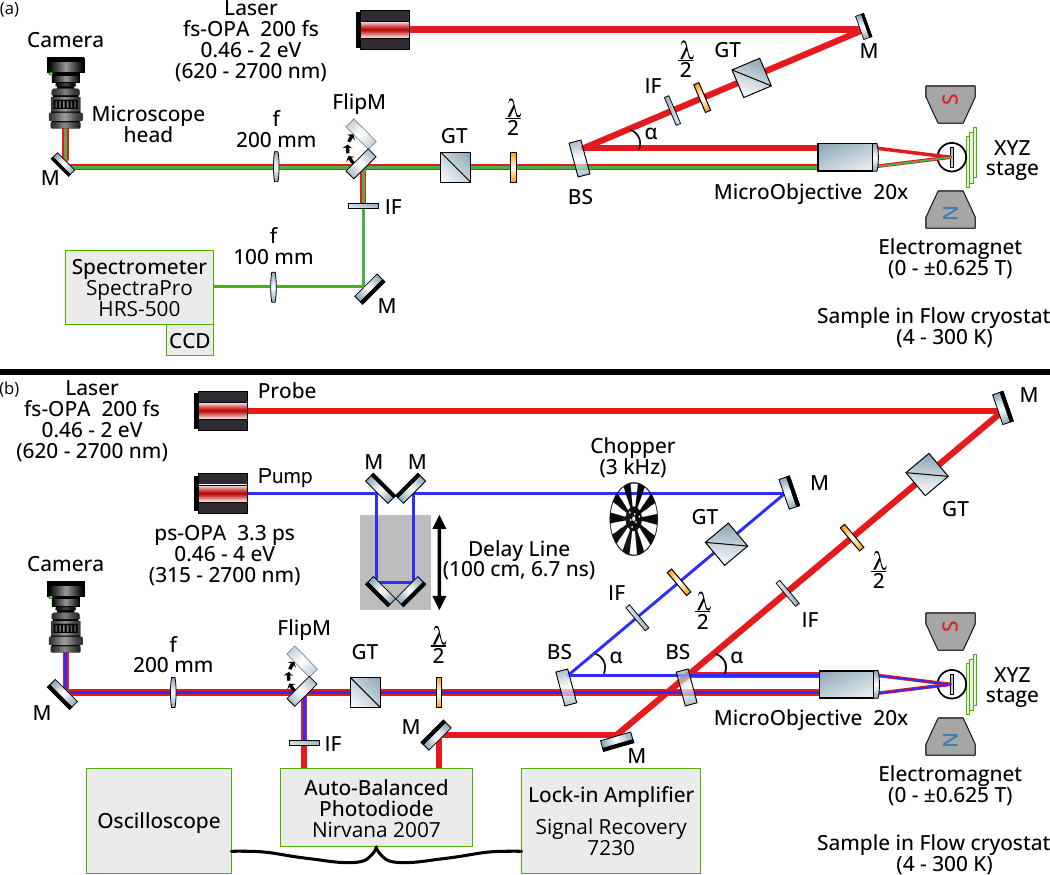}
\caption{Schematic representation of the experimental setup. (a) Confocal SHG microscope: the fs-OPA emission (200 fs, $0.46-2$\,eV, $620 - 2700$\,nm) (red) is routed through a Glan-Thompson polarizer (GT) and a half-wave plate ($\lambda/2$), reflected by a wedged beam splitter at small angle of $\alpha\approx8^\circ$, and focused by a 20× Mitutoyo Plan Apo NIR objective onto the sample in a helium flow cryostat ($4-300$\,K), where it is mounted on the three-axis linear stage (XYZ). The cryostat with the sample is placed between the poles of an electromagnet (±0.625 T). The back‐reflected SHG emission is analyzed by a second half‐wave plate and a Glan–Thompson polarizer; the flip‐mirror directs the beam either straight to the USB camera for real‐time visualization or, when engaged, reflects it through the interference filter into the SpectraPro HRS-500 spectrometer with a liquid nitrogen-cooled CCD for spectral acquisition attached.
(b) Two‐color pump-probe beam path: The Orpheus-fs (200 fs, $0.46-2$\,eV, $620-2700$\,nm) and Orpheus-ps (3.3 ps, $0.46-4$\,eV, $315-2700$\,nm) OPAs are both seeded by the Pharos amplifier and thus intrinsically synchronized. The pump beam is amplitude-modulated at 3 kHz using a mechanical chopper. The pump and probe beams are first conditioned by Glan-Thompson polarizers (GT) and half-wave plates ($\lambda/2$), then combined at a small angle of $\alpha\approx 8^\circ$ via wedged beam splitters (BS), and focused onto the sample by the objective. The wedged BS splits the incoming probe: the transmitted arm serves as the reference input to an auto-balanced photodiode (Nirvana 2007, PD), while the reflected arm continues to the sample. Light reflected by the sample retraces through the same wedges, and then passes through analyzer optics (GT, $\lambda/2$) to the signal input of the balanced photodiode. The differential output is demodulated by a lock-in amplifier (SR 7230) operated at a frequency referenced to the chopper. M is the label for a broadband mirror.
}
	\label{fig:optical_setup_scheme}

\end{figure*}

The SHG microscopy setup is built in a confocal back‐reflection geometry in which both the excitation and collection paths are coupled through a high‐NA, long‐working‐distance objective into a helium flow‐cryostat, see Fig.~\ref{fig:optical_setup_scheme}.

The laser system is a Light Conversion "Pharos" regenerative amplifier emitting 200 fs light pulses at 1030 nm wavelength (up to 9 W average power, up to 30 kHz repetition rate). This beam pumps two synchronized optical parametric amplifiers (OPAs): an Orpheus-ps OPA ($0.46-4$\,eV output photon energy, 3.3 ps pulse duration) and an Orpheus-fs OPA ($0.46-2$\,eV output photon energy, 200 fs pulse duration).

\subsection{Laser Beam Paths and Polarization Control}

The configuration of the setup for optical harmonic generation measurements is shown in Fig.~\ref{fig:optical_setup_scheme}(a). 
Short laser pulses are delivered by either a fs- or a ps-OPA, depending on the experimental requirements. The output beam is directed through a set of polarization optics components, consisting of a Glan-Thompson prism (GT) serving as linear polarizer and a motorized achromatic half-wave plate ($\lambda/2$), which enables a change of the linear polarization direction of the incoming light.

The beam is then guided by silver mirrors (M) and reflected by a beam splitter at a small angle of about $8^\circ$ into the optical axis of the microscope objective, which minimizes the polarization distortion of the s- and p-components upon reflection, avoiding the beam polarization to become elliptic. The beam is focused by a long-working-distance, 20$\times$ near-infrared objective (NA = 0.40) to a spot of about 2\,$\mu$m diameter on the sample surface.

The sample is mounted on the cold finger of a helium flow cryostat (temperature range $4-300$\,K), positioned between the poles of a water-cooled electromagnet, providing an in-plane magnetic field up to $\pm$0.625\,T strength. The entire cryostat is mounted on a three-axis translation stage (XYZ stage), allowing spatial scanning of the sample surface relative to the fixed focused laser spot.

The reflected excitation beam and the nonlinear optical signals (e.g., SHG) are collected by the same objective for collimation back along the incident path. After passing through another motorized half-wave plate and another Glan-Thompson prism, the polarization of the reflected signal can be analyzed in detection. A flip mirror (FlipM) allows switching between two detection modes: (i) \textbf{Imaging} mode, where the collimated beam is directed to a camera for live visualization of the illuminated sample region. To assist with alignment and targeting, a white-light lamp can be coupled into the incident path and focused onto the sample via the same objective. This enables identification of sample features (e.g., flakes) and accurate adjustment of the laser focus position for spectroscopy or microscopy scans. (ii) \textbf{Spectroscopy} mode, where the beam is directed into a 0.5\,m spectrometer (SpectraPro HRS-500). The beam is focused onto the spectrometer entrance slit and spectrally dispersed, after which it is detected by a nitrogen-cooled silicon CCD detector, enabling acquisition of nonlinear optical spectra with a resolution of about 60\,$\mu$eV.

The configuration for time-resolved pump-probe measurements is shown in Fig.~\ref{fig:optical_setup_scheme}(b). The ps-OPA provides the pump-pulses and the fs-OPA delivers the probe-pulses. The pump beam passes through a mechanical delay line with a length of 100\,cm, allowing a time delay of up to 6.7\,ns between the pump and probe pulses. A mechanical chopper modulates the pump beam at 3\,kHz frequency for lock-in detection.

The linear polarization angles of both beams can be independently adjusted using combinations of a Glan-Thompson prism and a motorized half-wave plate. Each beam is then reflected toward the microscope objective via its own beam splitter at a small incidence angle of about $8^\circ$, which minimizes polarization distortion upon reflection.

The beams are focused onto the sample through the microscope objective such that they spatially overlap. For this alignment, the imaging mode (i), as described above, can be used for optimizing the beams' focus and overlap on the sample surface.

The reflected probe beam is collected by the same objective and propagates back along the original probe path. After its polarization analysis, it is sent onto one of the detectors ('signal' channel) of an auto-balanced photodiode (Nirvana 2007), while a reference beam, namely a fraction of the probe split off ahead of the sample is delivered to the second photodiode ('reference' channel). The differential output is sent to a lock-in amplifier (Signal Recovery 7230) synchronized with the pump beam modulation frequency. This configuration allows for highly sensitive, time-resolved detection of weak nonlinear optical signals.

\subsection{Cryostat and Sample Environment}
The sample is placed in an Oxford "MicrostatHe-R" helium‐flow cryostat operating in the sample-in-vacuum configuration. To ensure optimal thermal contact, the samples are glued directly onto a copper cold finger using a vacuum-compatible adhesive. This cold finger resides within a sealed vacuum chamber maintained at a pressure of approximately \(1 \times 10^{-6}\,\mathrm{hPa}\). Such a high-vacuum environment minimizes the convective heat transfer and the sample contamination, thereby enabling a stable low-temperature operation and preserving the optical quality of sensitive nanostructures during extended measurements.
Its narrow, rectangular stainless‐steel tail ($30\times22$~mm) allows tight coupling to the microscope objective, while maintaining vacuum integrity. The temperature can be varied between 4\,K and 300\,K; with an additional resistive heater, temperatures up to 400\,K can be reached. The helium consumption is held below 0.5 L/h. A three‐axis XYZ linear stage (Edmund Optics Stock~\#66-460) accurately positions the cryostat with the sample. A water‐cooled electromagnet (GMW 5403 EG-50) placed just outside the cryostat tail can provide a magnetic field up to 0.625\,T parallel to the sample plane, i.e., in the Voigt configuration.

\subsection{Microscope and Detection}

A Thorlabs USB camera (DCC1545M-GL) is placed on top of the Zoom Inspection Microscope Head (Edmund Optics Stock~\#55-150) stand to provide an overhead black-and-white image of the sample surface, which is used for alignment.

The Mitutoyo Plan Apo NIR 20× objective (infinity‐corrected, 0.40 NA) focuses the excitation beams onto the sample and collects the emitted or reflected signals. After the flip mirror, the SpectraPro HRS-500 spectrometer disperses the light, which is then detected by a nitrogen-cooled silicon CCD (2048 × 512, 13 $\mu$m pixels); a typical spectral resolution under 0.2\,meV is achieved when using a 1200 grooves/mm grating. The best spectral resolution of 60\,$\mu$eV is achieved with a 2400 grooves/mm grating. The CCD signal integration times range from 100 ms (for strong SHG signals) to several seconds (for weak pump-probe transients).

\subsection{Optical Components and Filters}

\begin{itemize}
    \item \textbf{Polarizers and Retarders.}
Two Glan-Thompson polarizers (Thorlabs, GTH10M) serve as extinction elements: GT\,1 in the pump arm (transmission filter for the 1030 nm wavelength of the Orpheus‐ps), and GT\,2 in the probe arm (transmission filter for the Orpheus‐fs output). Achromatic $\lambda/2$ plates (B‐Halle, RAC 6.2.10L) on motorized OWIS rotation mounts allow automated polarization rotation of both the pump and probe beams. All polarizing elements have extinction ratios larger than $10^{4}\!:\!1$.
    \item \textbf{Beam Splitters.}  
UV Fused Silica Wedged Windows with 30 arcmin wedge angle (Thorlabs, WW41050) are oriented for a reflection angle of $\alpha \approx 8^\circ$, to minimize any polarization distortion over the spectral range of $400-1800$~nm.
    \item \textbf{Optical Filters.}
In the excitation path, the short‐pass interference filter (IF) (Thorlabs, FESHxxxx) was specifically selected for each targeted spectral range and installed after the half-wave plates, to block undesired small SHG signals generated in them.
In the collection path, a long‐pass IF (Thorlabs, FELHxxxx) rejects the fundamental light, while it lets the SHG light pass.
    \item \textbf{Mirrors.}
Protected silver mirrors (Thorlabs, Edmund Optics) handle reflection across the full experimental range of interest.
\end{itemize}

All optical components reside on a customized aluminum breadboard (Edmund Optics) with kinematic mounts to guarantee repeatable realignment following cryostat servicing or OPA maintenance. The near‐normal‐incidence arrangement of the mirrors significantly reduces polarization rotation ($\Delta\phi < 0.5^\circ$ over weeks of use), a feature that improves both the SHG signal fidelity and the pump-probe temporal overlap stability.

\begin{figure*}[!t]
	\begin{center}
        \includegraphics[width=0.95\textwidth]{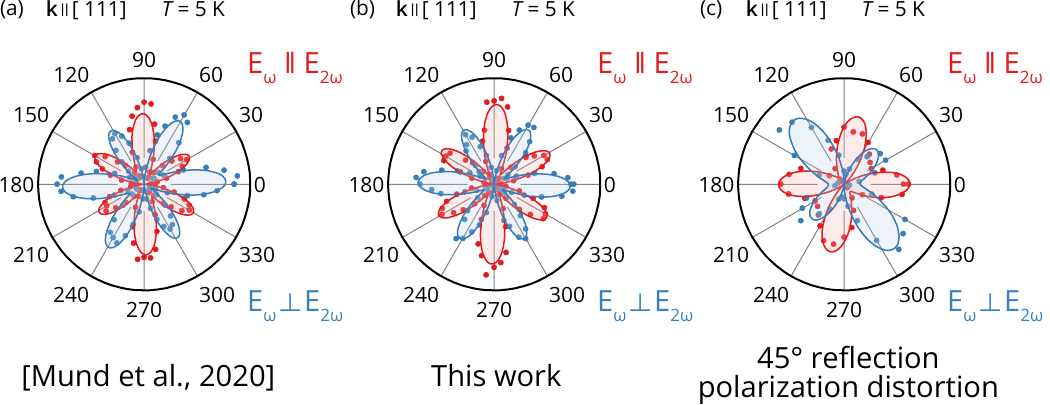}
\caption{SHG rotational‐anisotropy of the $1S$ exciton in bulk ZnSe [111] for three measurement settings.  
  (a) Reference data recorded in a liquid‐helium bath cryostat in transmission geometry~\cite{Mund20-ZnSe}.  
  (b) Data recorded using our confocal nonlinear multi-photon microscope (this work) with near‐normal‐incidence reflection on wedges. The excellent agreement with the reference data confirms the correct system performance, in particular with respect to the polarization control.
  (c) Data recorded on the same setup, where a dichroic mirror was used for reflection at 45° incidence, introducing polarization ellipticity and distorting the SHG lobes.  
 In all panels, the red symbols and shaded lobes correspond to the parallel \textbf{E}$_{\omega} \parallel$ \textbf{E}$_{2\omega}$ configuration, while the blue symbols and shaded areas correspond to the crossed \textbf{E}$_{\omega} \perp$ \textbf{E}$_{2\omega}$ configuration.
}
		\label{fig:ZnSe_111_RA}
	\end{center}
\end{figure*}

\subsection{Software and Automation}
All hardware components are controlled via a customized MATLAB (R2025a) graphical user interface developed in App Designer.  Key features comprise:
\begin{itemize}
    \item \textbf{Laser Tuning.} The Orpheus‐fs and Orpheus‐ps OPAs are tuned via the TOPAS API (Light Conversion). The photon energy setpoints and crystal angles are accessed through MATLAB functions calling the vendor's dll.
    \item \textbf{Polarization and Power Control.}  Motorized half‐wave plates (B. Halle $\lambda/2$) are mounted on OWIS rotational stages. The rotation angles (both for pump and probe) are driven by the OWIS controller via provided dll files, providing automated adjustment of beam power and linear polarization.
    \item \textbf{Delay Line.} The 100 cm Newport linear stage (delay line) is operated through the OWIS controller. MATLAB scripts move the stage in 10 $\mu$m steps ($\approx$67 fs delay) under program control.
    \item \textbf{Spectrometer and CCD.}  The Princeton Instruments SpectraPro HRS‐500 spectrometer and the nitrogen-cooled silicon CCD are accessed via the LightField Add-in and Automation SDK. Spectral acquisition and integration times are managed by MATLAB routines calling LightField API objects.
    \item \textbf{Powermeter.}  A Thorlabs PM100D thermopile powermeter head communicated via USB. MATLAB addressed Thorlabs' drivers (PM100D.dll) to query real‐time power readings during alignment and data collection.
    \item \textbf{Flip‐Mirror Control.} Thorlabs motorized flip‐mirrors (FPM100) in both the pump and probe paths are toggled via serial (COM‐port) ASCII commands from MATLAB, enabling rapid switching between alignment and measurement configurations.
\end{itemize}
All spectral data are initially saved by LightField as plain‐text ".csv" files; a parallel MATLAB routine converted these to raw ".bin"format on‐the‐fly for accelerated reading and downstream analysis.

\section{Experimental test}

In this section, we experimentally demonstrate the potential of the confocal microscopy setup for various nonlinear optical experiments. To show the versatility, precision, and sensitivity of the system, we perform measurements on well-characterized materials and compare the results to previous data obtained using a conventional nonlinear optical setup with a bath cryostat, additionally demonstrating the extensions that go beyond such a system, like the possibility to study micrometer-sized flake samples.
First, we verify the polarization control in both the excitation and detection beam paths by performing polarization tomography on the $1S$excitons in ZnSe and Cu$_2$O bulk crystals. 
Next, we demonstrate the ability to apply a magnetic field in Voigt geometry by demonstrating the magnetic-field-induced SHG contributions that distort the polarization diagrams. 
The broad spectral tuning range of the setup is showcased by a SHG spectral scan from 1.4 to 3.1\,eV in ZnSe. 
We further illustrate the spatial mapping possibility by imaging different stacked and twisted layer configurations of MoS$_2$ flakes. 
Finally, we highlight the potential for time-resolved studies by performing two-color pump–probe measurements of coherent phonons in perovskites.

\subsection{Polarization tomography}
\begin{figure}[t]
	\begin{center}
        \includegraphics[width=0.99\columnwidth]{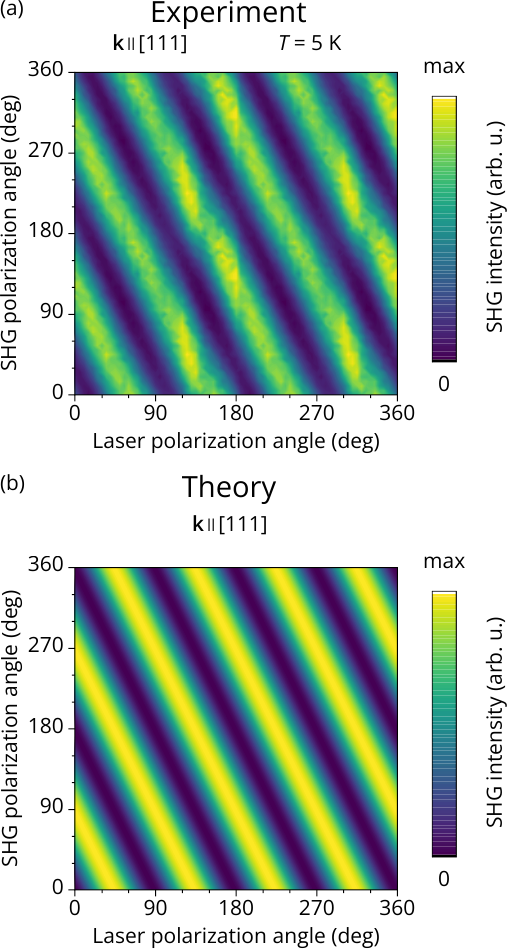}
		\caption{Polarization-resolved (SHG) in Cu$_2$O  for light propagating along the [111] direction, showing both experiment and theory on the same color scale (normalized to the maximum in each panel).
    (a) Confocal SHG measurement on the 1\textit{S} exciton at $T=5\,$K: the incident laser polarization is rotated through $0-360$° by a $\lambda/2$ plate (horizontal axis), and the SHG signal polarization is analyzed by a second $\lambda/2$ plate (vertical axis).
    (b) Corresponding group‐theory calculation of the SHG intensity in Cu$_2$O.
		}
		\label{fig:Cu2O_tomography}
	\end{center}
\end{figure}
Here, we demonstrate the independent control of the linear polarization angles in both the excitation and detection paths, without distortion of the s- and p-polarization components. A common method in SHG studies is the measurement of rotational anisotropies, where the SHG intensity is recorded as function of the linear polarization angle $\psi$ of the incident laser light, which is tuned simultaneously with the linear polarization angle $\varphi$ of the SHG signal in the detection path. The angles are tuned from 0$^\circ$to 360$^\circ$ for the parallel ($\psi=\varphi$) and crossed ($\varphi=\psi+90^\circ$) configurations.

The rotational anisotropies of the $1S$ exciton in ZnSe are well documented and previously were measured at $T = 5$\,K for $\mathbf{k} \parallel [111]$ in a conventional SHG setup using a bath cryostat~\cite{Mund20-ZnSe}, as shown in Fig.~\ref{fig:ZnSe_111_RA}(a). Corresponding experimental data measured in angle steps of 5$^\circ$ are shown by the dots. The parallel \textbf{E}$_{\omega} \parallel$ \textbf{E}$_{2\omega}$ (red) and crossed \textbf{E}$_{\omega} \perp$ \textbf{E}$_{2\omega}$ (blue) configuration data show sixfold patterns, which are characteristic for this sample orientation. Here, \textbf{E}$_{\omega}$ and \textbf{E}$_{2\omega}$ denote the electric field vectors of the fundamental and second-harmonic light, respectively.

For comparison, Fig.~\ref{fig:ZnSe_111_RA}(b) displays the rotational anisotropy obtained for the otherwise identical experimental conditions using our confocal microscopy setup, where the incident laser light is reflected under a small angle after passing through the polarization optics. The results are in good agreement with the literature reference data, confirming the high quality of the rotational anisotropy detection.

By contrast, Fig.~\ref{fig:ZnSe_111_RA}(c) presents the measurement data obtained using a $45^\circ$ reflection behind the polarization optics. This geometry introduces considerable distortions of the s- and p-polarization components, resulting in elliptical polarization and a loss of the characteristic sixfold symmetry.

To better understand the underlying SHG mechanisms and the symmetries of the exciton states, it is useful to go beyond the standard measurements in just the parallel and crossed configurations and instead record the SHG intensity for all possible combinations of the polarization angles $\psi$ and $\varphi$~\cite{Farenbruch20,Farenbruch2021}. To that end, an automated measurement protocol is applied for the motorized half-wave plates. The excitation polarization angle $\psi$ is fixed, and the detection angle $\varphi$ is scanned from $0^\circ$ to $360^\circ$. Then, $\psi$ is increased by $10^\circ$, and the scan of $\varphi$ is repeated. This is continued until the full $(\psi, \varphi)$ space is covered. With an integration time of one second per spectrum, a complete polarization tomography map can be acquired in about one hour. The SHG intensity is plotted as a function of $\psi$ and $\varphi$ in a two-dimensional color map in Figure~\ref{fig:Cu2O_tomography}(a), which shows the result for the $1S$ exciton in Cu$_2$O ($\mathbf k\parallel [111]$), using scanning steps of $10^\circ$. Diagonal stripe patterns are observed, resulting in a sixfold symmetry for the parallel and crossed polarization configurations, while it leads to fourfold symmetry for fixed $\varphi$ and twofold symmetry for fixed $\psi$.

The corresponding theoretical calculation is shown in Fig.~\ref{fig:Cu2O_tomography}(b). The modeling is based on a group-theoretical symmetry analysis, as discussed in detail for crystallographic and magnetic-field-induced SHG of bright and dark excitons in Refs.~\cite{Farenbruch2020PRL,Farenbruch20}. The tomography measurement provides distinct patterns, which help one to identify the exciton symmetries and nonlinear optical mechanisms. It also allows identification of polarization configurations where specific mechanisms dominate, while others are suppressed. This was demonstrated in Ref.~\cite{Farenbruch2025arXiv} for state-selective control of exciton quantum beats.

\subsection{Experiments in magnetic field}

To demonstrate the application potential of an external magnetic field using the electromagnet (for reaching $B = 0.625$\,T maximum in Voigt geometry), we select a bulk Cu$_2$O crystal with the incident laser beam directed along the $[1\bar{1}0]$ crystallographic axis. Although this crystal orientation is symmetry-forbidden for SHG, the triply degenerate $1S$ exciton becomes allowed due to $k^2$ band structure-induced and strain-induced splitting effects, which lift the degeneracy by a few~$\mu$eV~\cite{Mund2019-PRB}.

\begin{figure}[!t]
	\begin{center}
         \includegraphics[width=0.99\columnwidth]{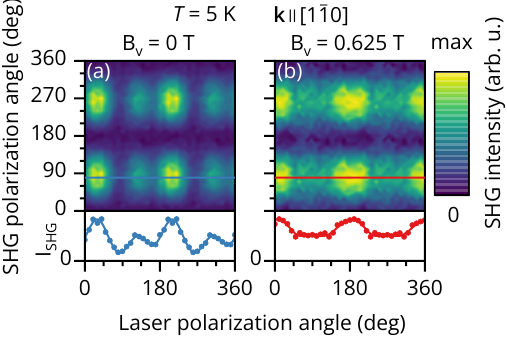}       
		\caption{Polarization‐resolved SHG tomography of bulk Cu$_2$O (\textbf{k} $\parallel[1\bar{1}0]$) at $T=5\,$K at zero magnetic field (a) and $B=0.625\,$T (b). The color maps display the SHG intensity as a function of the linear polarization angle of the fundamental excitation ($\omega$, horizontal axis) and of the second harmonic emission ($2\omega$, vertical axis). The bottom panels show representative line cuts at fixed SHG analyzer angle with the fundamental polarization scanned: blue curve for $B=0\,$T and red curve for $B=0.625\,$T. The applied magnetic field modifies the SHG symmetry - reducing the number of lobes, shifting the angular positions of the maxima, and lifting the zero‐intensity nodes observed at zero field, thereby demonstrating the setup sensitivity to magnetic field‐induced SHG mechanisms.
		}
		\label{fig:Cu2O_magnetic_field}
	\end{center}
\end{figure}

At zero magnetic field, the polarization tomography map exhibits a fourfold symmetry for fixed $\varphi$ and a twofold symmetry for fixed $\psi$, as shown in Fig.~\ref{fig:Cu2O_magnetic_field}(a). When a magnetic field of $B = 0.625$\,T is applied in Voigt geometry, the tomography map changes noticeably, as can be seen in Fig.~\ref{fig:Cu2O_magnetic_field}(b): the pattern becomes distorted due to additional magnetic-field-induced SHG contributions, which interfere with those arising from the $k^2$ and strain mechanisms. While the twofold symmetry for fixed $\psi$ remains, the dependence for fixed $\varphi$ changes to a twofold pattern with a non-vanishing background.

\subsection{Spectral tunability}

\begin{figure}[!htbp]
	\begin{center}
        \includegraphics[width=0.99\columnwidth]{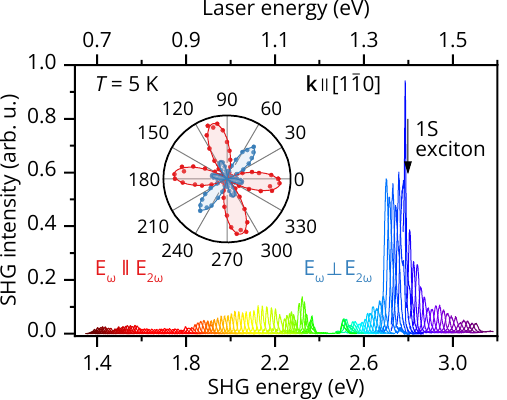}
		\caption{SHG intensity measured from a bulk ZnSe sample with propagation vector \textbf{k} $\parallel[1\bar{1}0]$ at $T=5\,$K, plotted versus SHG photon energy (lower axis) and fundamental photon energy (upper axis).  The spectrum covers a wide range from 1.4 to 3.1 eV. The spectra around 2.4 eV are not displayed because the ps-OPA switched from signal to idler mode, causing issues with the filter optics described in main text.
    Rotational anisotropies of the SHG signal at the 1S exciton resonance, recorded in the parallel \textbf{E}$_{\omega} \parallel$ \textbf{E}$_{2\omega}$ and crossed \textbf{E}$_{\omega} \perp$ \textbf{E}$_{2\omega}$ polarization geometries, are shown in the inset.
    Together, these data demonstrate the broad spectral tunability and full polarization-analysis potential of the confocal microscopy SHG setup.
        }
		\label{fig:ZnSe_shg_scan}
        \end{center}
\end{figure}

An SHG scan of a bulk ZnSe crystal at $T = 5$\,K, with the incident laser beam propagating along the $[1\bar{1}0]$ crystallographic direction, is shown in Fig.~\ref{fig:ZnSe_shg_scan}. The linear polarizations of the fundamental and SHG light were both aligned parallel to the $[111]$ direction. The plot displays individual SHG spectra recorded with picosecond laser pulses for fundamental photon energies tuned from 0.70 to 1.55~eV, corresponding to a SHG energy range of $1.4-3.1$\,eV.  

For photon energies below 2.70~eV, the SHG signal remains comparatively weak.  
The spectra around 2.4\,eV are not shown in the diagram, as the signal and idler output beams of the OPA are spectrally too close to be separated with standard FESH/FELH filters, resulting in overlapping SHG contributions from both beams. A practical solution is to insert an additional Glan-Thompson polarizer after the OPA output.
Above 2.70\,eV, the SHG response increases significantly and exhibits a sharp maximum at 2.80~eV, which we assign to the $1S$ exciton resonance. At higher photon energies above the band gap, the SHG intensity decreases steadily, approaching a low level at 3.10~eV.

The inset shows the SHG rotational anisotropy for parallel (red) and crossed (blue) polarization configurations. The parallel configuration reveals a fourfold symmetry, whereas the crossed configuration exhibits a sixfold pattern with two dominant "leaves" at about $50^\circ$.  

These measurements demonstrate the broad spectral tunability of the confocal SHG setup, which in principle can be extended to a spectral SHG range from 0.8 to 7.8\,eV.

\subsection{Spatial microscopy mapping}

\begin{figure*}[hbt]
	\begin{center}
        \includegraphics[width=0.99\textwidth]{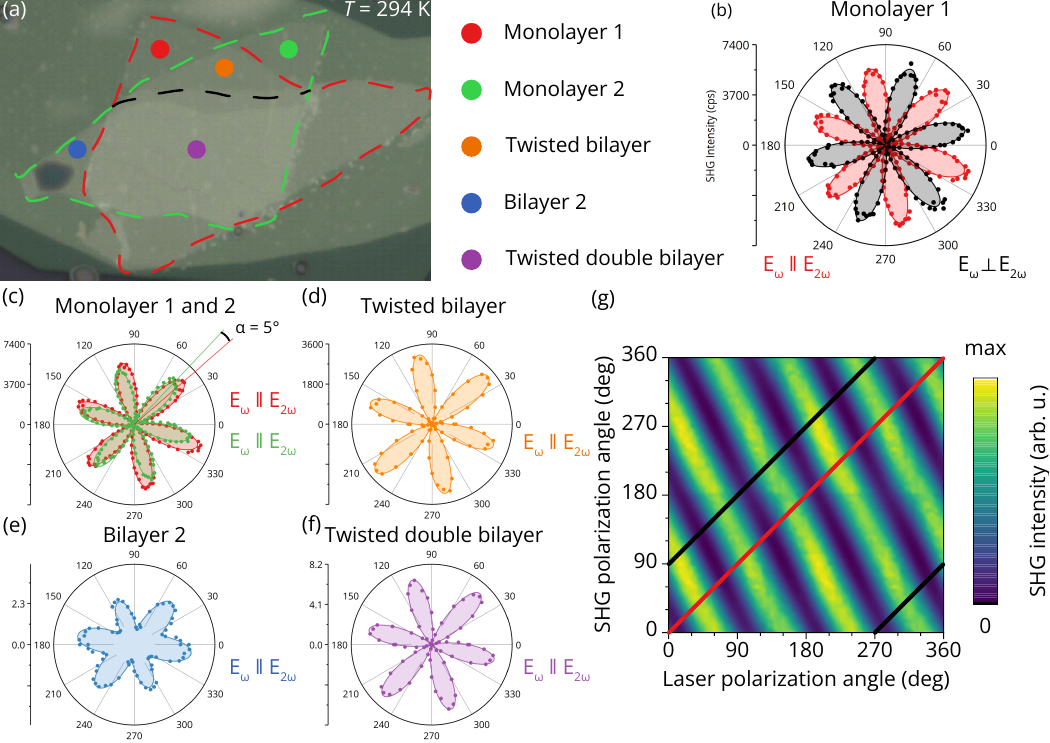}
		\caption{SHG confocal microscopy and rotational‐anisotropy measurements on MoS$_2$ flakes.  
(a) Bright‐field optical micrograph of the MoS$_2$ sample with two flakes placed one (red dashed contour) on top of the other (green dashed contour) with a twist angle. Colored dots mark the locations where rotational‐anisotropy scans were recorded in the following areas: first monolayer (red), second monolayer (green), twisted bilayer (orange),  bilayer (blue), and twisted double bilayer (purple).
(b) Rotational SHG anisotropy measured in the first monolayer region. Red and black dots represent data for parallel \textbf{E}$_{\omega} \parallel$ \textbf{E}$_{2\omega}$ and crossed \textbf{E}$_{\omega} \perp$ \textbf{E}$_{2\omega}$ configuration, respectively.
(c–f) Rotational anisotropies of the measured SHG intensity in parallel \textbf{E}$_{\omega} \parallel$ \textbf{E}$_{2\omega}$ configuration for each region: (c) overlay of monolayer 1 (red) and monolayer 2 (green) illustrating a relative twist angle of $\approx5°$, (d) twisted bilayer, (e) bilayer 2, (f) twisted double bilayer. Solid lines are fits to the expected sixfold SHG symmetry.
(g) Two‑dimensional color map of SHG intensity comprising all possible combinations of linear polarizations of incoming and outgoing light for the monolayer 1 region. The parallel and crossed configurations are indicated as red and black diagonal lines.
		}
		\label{fig:MoS2_mapping}
	\end{center}
\end{figure*}

To demonstrate the microscopy potential, we studied a mechanically exfoliated MoS\textsubscript{2} flake in which two monolayers were stacked with a twist angle of $\approx5^\circ$, producing adjacent regions of monolayer, bilayer, twisted bilayer and twisted double bilayer. Figure~\ref{fig:MoS2_mapping}(a) shows a bright‐field micrograph with color‐coded contours. The dots indicate the positions where we collect the SHG anisotropy scans: monolayer 1 (red), monolayer 2 (green), twisted bilayer (orange), bilayer (blue) and twisted double bilayer (purple).
The rotational anisotropy from monolayer 1, measured at 1.44\,eV excitation photon energy (below the band gap, but above the C‐exciton), in both parallel (\textbf{E}$_{\omega}\parallel \textbf{E}_{2\omega}$, red) and crossed ($\textbf{E}_{\omega}\perp \textbf{E}_{2\omega}$, black) polarization configuration, is shown in Fig.~\ref{fig:MoS2_mapping}(b).
The six‐lobe SHG patterns for each region in parallel geometry are shown in the other panels of Fig.~\ref{fig:MoS2_mapping}: monolayer 1 (c), overlay of two twisted monolayers with a $5^\circ$ twist (d), bilayer with suppressed signal (e) and twisted double bilayer (f).
The full 2D polarization map for monolayer 1 in panel (g), indicating also cuts for the parallel (red diagonal) and crossed (black diagonal) polarization configuration, highlights the possibility for sensitive detection of any polarization‐dependent distortions.

\noindent\textbf{Monolayer SHG and Crystal Orientation.}  In monolayer 1, we observe a pronounced sixfold SHG pattern, confirming the $D_{3h}$ symmetry of single‐layer MoS\textsubscript{2}. The orientations of the six lobes in red and black [Fig.~\ref{fig:MoS2_mapping}(b)] allow us to accurately determine the crystal axes with sub-degree precision. Repeating the measurement on monolayer 2 and overlaying the two curves [Fig.~\ref{fig:MoS2_mapping}(c)] directly yields the twist angle of $\approx5^\circ$ between them~\cite{Hsu2014ACSNano,Paradisanos2022PRB}.

\noindent\textbf{Bilayer versus\ Twisted Bilayer.} In the AB‐stacked bilayer region, inversion symmetry nominally forbids SHG; we detect only a weak residual signal [Fig.~\ref{fig:MoS2_mapping}(e)], indicating the weak influence of stacking faults or strain.  In contrast, the twisted bilayer restores non-centrosymmetry via the moiré superlattice, producing a clear sixfold SHG pattern of intermediate intensity [Fig.~\ref{fig:MoS2_mapping}(d)].

\noindent\textbf{Twisted Double Bilayer.}  Finally, in the region where two AB bilayers are stacked with a twist, we again detect SHG [Fig.~\ref{fig:MoS2_mapping}(f)].  Although each bilayer alone is centrosymmetric, the overlapping twist creates local AA‐stacking domains that generate SHG, even though with lower amplitude than the monolayer.

\noindent\textbf{Polarization Tomography.} The 2D map in Fig.~\ref{fig:MoS2_mapping}(g) demonstrates our ability to sweep both the input and analyzer angles, reconstructing the full polarization dependence of the SHG. The technique can detect subtle ellipticity or field‐induced symmetry breaking and will be valuable for future studies in magnetic or electric fields.

In summary, the confocal microscopy SHG setup provides: (i) micrometer‐scale mapping of SHG intensity across complex 2D heterostructures, (ii) high‐precision determination of crystal orientation and twist angles via rotational anisotropies, and (iii)
comprehensive polarization tomography for identifying the symmetry of electronic states.
Although all measurements here are made at room temperature, our flow‐cryostat enables experiments down to 4\,K. Looking forward, we plan to combine SHG spectroscopy with two‐color pump–probe and magneto‐optical measurements to explore exciton and many‐body phenomena in twisted van der Waals materials.

\subsection{Two-color pump-probe}

To demonstrate the time-domain potential of the microscope setup, we implemented pump--probe Brillouin spectroscopy on bulk Cs$_2$AgBiBr$_6$ with \textbf{k} $\parallel[111]$, where shear phonons are launched via giant anisotropic photostriction~\cite{Horiachyi25}. 
The optical setup is shown in Fig.~\ref{fig:optical_setup_scheme}(b): the 3.3~ps pump at 2.81\,eV photon energy and the time-delayed probe at 2\,eV photon energy are taken from the synchronized ps- and fs-OPAs and co-focused through the same 20$\times$, NA~0.40 Mitutoyo objective. The pump pulses are chopped at 3~kHz frequency. The reflected probe is detected with an auto-balanced Si photodiode using lock-in detection. The mechanical delay line covers time delays between the pump and probe pulses up to 6.7~ns. The average powers at the sample are kept below 20~$\mu$W to minimize sample heating. The measurements are performed at 30~kHz repetition rate.

\begin{figure*}[hbt]
	\begin{center}
        \includegraphics[width=0.99\textwidth]{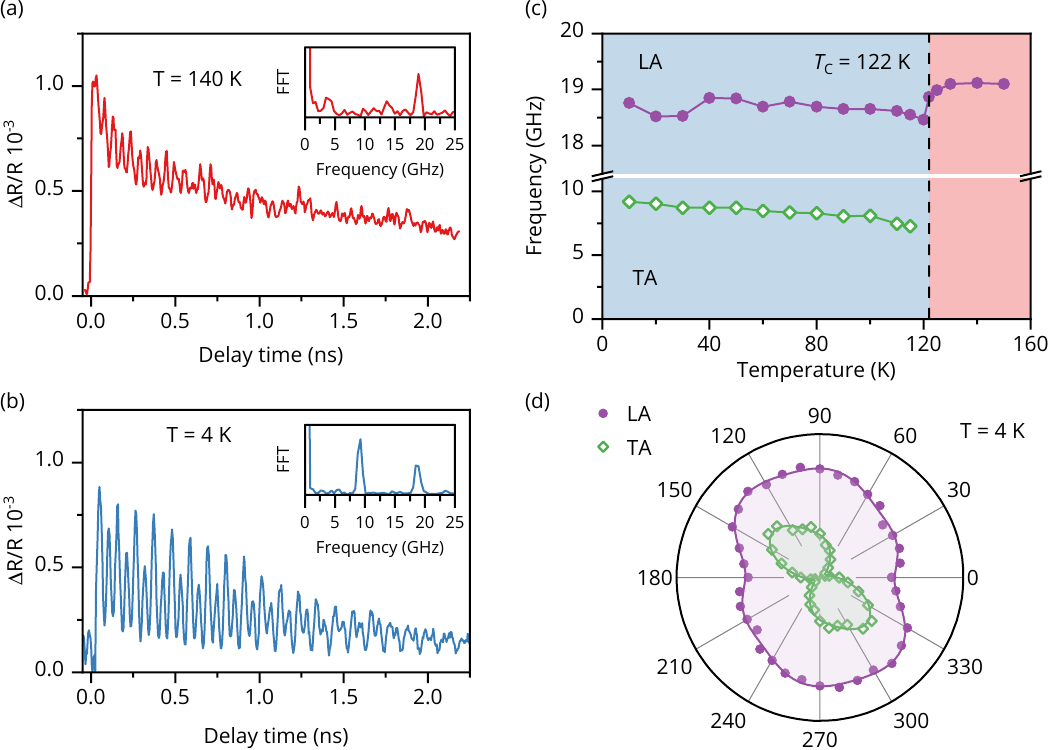}
		\caption{
Two-color pump--probe of Cs$_2$AgBiBr$_6$.
(a) Transient differential reflectivity $\Delta R/R$ recorded above the structural phase transition temperature at 140\,K. The inset shows the fast Fourier transform (FFT) with a single peak at $\sim 19$\,GHz, assigned to the longitudinal acoustic (LA) mode; the transverse (TA) branch is symmetry-forbidden in the cubic phase.
(b) $\Delta R/R$ trace well below $T_\mathrm{C}$ at 4\,K, with the FFT given in the inset. The two distinct peaks---TA at $\sim 9$\,GHz and LA at $\sim 19$\,GHz---demonstrate simultaneous detection of shear and longitudinal phonons.
(c) Temperature evolution of the Brillouin frequencies: purple circles denote the LA phonons and green diamonds denote the TA phonons. The LA branch shows a clear kink at $T_\mathrm{C}$, marking the structural transition, while the TA mode softens and disappears as $T_\mathrm{C}$ is approached from below.
(d) Rotational-anisotropy polar plot of the Brillouin amplitude at 4\,K, measured while rotating the linear polarization of both the incident and analyzed probe beams (parallel geometry, identical to the SHG rotational anisotropy). Purple circles: LA branch; green diamonds: TA branch. The two-lobed TA pattern indicates the strong anisotropic photostriction as the dominant driving mechanism, whereas the LA response is nearly isotropic with a slight elongation along the $c$ axis.
All measurements were performed at 30\,kHz repetition rate; the $\sim 33$\,$\mu$s interpulse interval, together with low average powers, prevents cumulative laser heating.
}
		\label{fig:CABB_pump-probe}
	\end{center}
\end{figure*}

Representative transients are given in Figs.~\ref{fig:CABB_pump-probe}(a,b). In the cubic phase of Cs$_2$AgBiBr$_6$ at $T = 140$\,K, the differential reflectivity $\Delta R/R$ exhibits a single oscillatory component. Accordingly, its Fast Fourier Transform (FFT) spectrum shows one peak at $\approx 19$~GHz, assigned to the longitudinal acoustic (LA) mode. The transverse acoustic (TA) mode is symmetry-forbidden in the cubic phase. Cooling well below the orthorhombic phase transition ($T_\mathrm{C} = 122$\,K) activates shear phonon dynamics: at $T = 4$\,K two well-resolved frequencies emerge, the TA mode at $\approx 9$~GHz and the LA mode at $\approx 19$\,GHz, as seen directly in the temporal trace and in the FFT spectrum in Fig.~\ref{fig:CABB_pump-probe}(b). 

Figure~\ref{fig:CABB_pump-probe}(c) summarizes the temperature evolution of the phonon modes. The LA branch (purple circles) exhibits a clear kink at $T_\mathrm{C}$, marking the structural transition, while the TA branch (green diamonds) softens with increasing temperature and vanishes within a few Kelvin around $T_\mathrm{C}$. This temperature-dependent visibility of the TA mode and the kink for the LA mode provides a convenient thermometer for phase identification. The low average power at 30~kHz pulse repetition rate ($\approx 33~\mu$s repetition period) are essential here: they prevent heating and allow the lattice to reach equilibrium between pulses, so that the intrinsic transition is resolved without thermal smearing.

To identify the driving mechanism, we measure the rotational-anisotropy of the phonon amplitude at $T = 4$\,K. The TA response follows a pronounced two-lobe pattern consistent with a $\cos 2\phi$ dependence, indicative of anisotropic photostriction as the dominant source of shear-mode excitation. In contrast, the LA amplitude is nearly isotropic with only a weak elongation along the crystallographic $c$ axis [Fig.~\ref{fig:CABB_pump-probe}(d)].

These results highlight two practical advantages of our implementation. First, the microscope-based, low-repetition-rate module retains the frequency resolution associated with cw Brillouin light scattering while avoiding laser-induced heating, a limitation in continuous-wave studies. Second, the same optical setup that supports polarization-resolved SHG can, without big reconfiguration, be applied to time-resolved measurements of both LA and TA coherent phonons with full polarization control.

\section{Conclusions and Outlook}

We have implemented a confocal microscopy setup for nonlinear multi-photon spectroscopy that combines: 
\begin{enumerate}[label=(\roman*)]
    \item Full polarization control in both excitation and detection.
     \item Polarization-preserving beam illumination via near-normal incidence on wedged beam splitters (avoiding $s$ and $p$ distortion).
      \item Broadband spectral tunability with spectral analysis of the emitted harmonics.
       \item Diffraction-limited spatial scanning in a microscope geometry.
        \item Operation from cryogenic to room temperature.
        \item Compatibility with in-plane magnetic fields and time-resolved two-color pump-probe excitation with balanced detection.
\end{enumerate}
Together, these potentials enable quantitative rotational-anisotropy measurements and full polarization tomography of SHG and higher-order signals from bulk crystals and micron-scale nanostructures.

The approach is broadly applicable to any material whose optical transitions fall within the laser spectral range. It is particularly powerful under resonant excitation conditions, for example, near exciton resonances in semiconductors, where the SHG response is strongly enhanced and becomes highly symmetry-selective. The polarization-tomography backbone is designed for symmetry analysis and for tracking controlled symmetry modifications under variation of external parameters, including strain, gate-controlled electric fields, magnetic fields, and tailored photoexcitation.

Looking forward, the same platform can be extended to third- and fourth-harmonic generation (THG/FHG), multi-wave-mixing (SFG/DFG/FWM), and valley-selective measurements using circularly polarized excitation. These upgrades will enable systematic, resonant, symmetry-resolved nonlinear spectroscopy for systems as diverse as bulk semiconductors, van der Waals heterostructures, and other nanostructured materials within a single, versatile instrumental setup.

\section*{ACKNOWLEDGMENTS}
We thank D. O. Horiachyi and I. A. Akimov for fruitful discussions. We acknowledge the financial support by the Deutsche Forschungsgemeinschaft in the frame of the Collaboration Research Center TRR142 (project A11). X.H., D.J.G. and A.I.T. were supported by the EPSRC grants EP/V006975/1, EP/V026496/1. All authors acknowledge support by the EPSRC Centre-to-Centre grant EP/S030751/1.

\section*{AUTHOR DECLARATIONS}

\section*{Conflict of interest}
The authors have no conflicts to disclose.

\section*{Author Contributions}

\textbf{Nikita V. Siverin}: Investigation (lead), Methodology (lead), Writing - original draft (leading), Writing - review \& editing (equal),  
\textbf{Andreas Farenbruch}: Investigation (equal), Writing - review \& editing (equal).
\textbf{Dmitri R. Yakovlev}: Conceptualization (lead), Investigation (equal),  Writing (equal), Supervision (lead).
\textbf{Daniel J. Gillard}:  Investigation (equal).
\textbf{Xuerong Hu}:  Investigation (equal).   
\textbf{Alexander I. Tartakovskii}: Conceptualization (lead),  Supervision (lead).
\textbf{Manfred Bayer}: Funding acquisition (lead), Writing - review \& editing (equal), Supervision (supporting).

\section*{DATA AVAILABILITY}

The data that support the findings of this study and software for automation and analysis are available from the corresponding author upon reasonable request.

\end{document}